\documentclass[fleqn,12pt,twoside]{article}
\usepackage{espcrc1}
\usepackage{epsf}

\usepackage{graphicx}
\usepackage[figuresright]{rotating}

\newcommand{\AmS}{{\protect\the\textfont2
  A\kern-.1667em\lower.5ex\hbox{M}\kern-.125emS}}

\title{Overview on All Reactions Linked to GPDs}

\author{Marc Vanderhaeghen \address[]{Institut f\"ur Kernphysik,
	Johannes Gutenberg-Universit\"at, 
	D-55099 Mainz, Germany}%
	}
       
\begin{document}

% typeset front matter
\maketitle

\begin{abstract}
A short overview is given on how generalized parton
distributions (GPDs) enter in a variety of hard exclusive processes such as 
deeply virtual Compton scattering (DVCS) and hard
meson electroproduction reactions on the nucleon.
We firstly discuss the links between GPDs and elastic nucleon form
factors which represent powerful constraints on parametrizations of GPDs.
Subsequently, we show some key observables which 
are sensitive to the various hadron structure aspects of the GPDs,  
and which are at present under experimental investigation 
at different facilities (HERMES, H1/ZEUS, JLab and Compass), 
or will be addressed by experiments in the near future.
\end{abstract}

\section{Introduction}

\begin{figure}[h]
\vspace{-10cm}
\epsfxsize=16 cm
\centerline{\epsffile{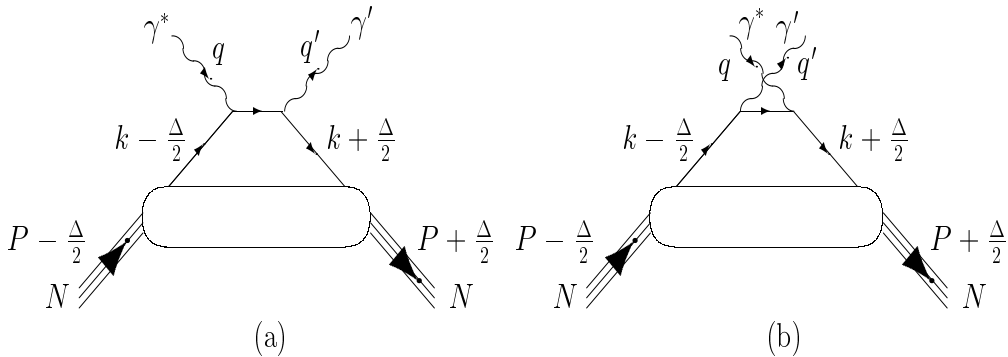}}
\vspace{-9.25cm}
\caption[]{\small ``Handbag'' diagrams for the DVCS process, 
containing the GPDs.}
\label{fig:handbags}
\end{figure}

Generalized parton distributions (GPDs), 
are universal non-perturbative objects entering the
description of hard exclusive electroproduction processes 
(see Refs.~\cite{Ji98,Rad01,GPV01} for reviews and references).
In leading twist there are four GPDs for the nucleon, 
i.e. $H$, $E$, $\tilde H$ and $\tilde E$, which 
are defined for each quark flavor ($u$, $d$, $s$). 
These GPDs depend upon the different longitudinal momentum fractions 
$x + \xi$ ($x - \xi$) of the initial (final) quark and upon the
overall momentum transfer $t = \Delta^2$ to the nucleon 
(see Fig.~\ref{fig:handbags}). 
As the momentum fractions of initial and final quarks are
different, one accesses quark momentum correlations in the
nucleon. Furthermore, if one of the quark momentum fractions is
negative, it represents an antiquark and consequently one may
investigate $q \bar q$ configurations in the nucleon. 
Therefore, these functions contain a wealth of new nucleon 
structure information, generalizing the information obtained in
inclusive deep inelastic scattering.  
\newline
\indent
To access this information, a general parametrization for all four
GPDs has been given in~\cite{GPV01}. 
For the GPDs $H$ and $E$, 
a two-component parametrization has been developed consisting
of a double distribution \cite{Mul94,Rad99} and a D-term \cite{Pol99} part. 
The latter is sensitive to scalar-isoscalar $q \bar q$ configurations in the
nucleon. 
The parameters entering this parametrization
can be related in a rather general way to such (not yet measured)
quantities as the contribution of the nucleon spin carried by the quark
total angular momentum ($J^u, J^d$, etc.),
$\bar q q$ components of the nucleon wave function
(in particular the D-term and the ``vector meson'' part of the GPD $E$),
the strength of the skewedness effects in the GPDs (encoded in their  
$\xi$-dependence),
the quark structure of $N \to N^*, \Delta$ transitions, 
the weak electricity GPD in nucleon to hyperon transitions,
flavor $SU(3)$ breaking effects, and others.
Furthermore, it has been shown that by a Fourier transform of the
$t$-dependence of GPDs, it is conceivable to access the
distributions of parton in the transverse plane, see Refs.~\cite{Bur00,Die02}. 
\newline
\indent
In this short paper, only a very limited selection of the above topics can be
touched upon, and the reader is referred to the other contributions in these
proceedings for more discussions on the above subjects. 

\section{Nucleon electromagnetic form factors}

\begin{figure}[h]
\vspace{-1.75cm}
\epsfxsize=10cm
\centerline{\epsffile{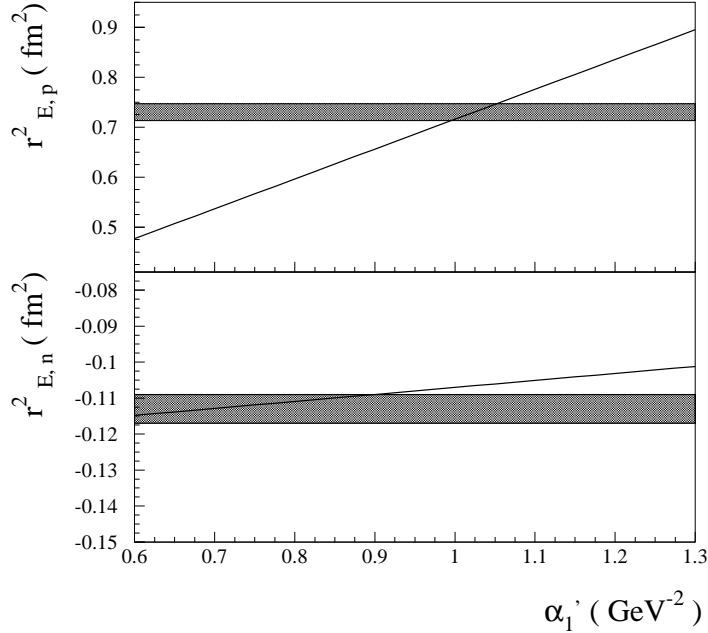}}
\vspace{-1cm}
\caption[]{\small Proton and neutron electric mean squared radii 
$r^2_{E, p}$ (upper panel) and $r^2_{E, n}$ (lower panel),   
Eq.~(\ref{eq:rmse}). 
The calculations show the dependence of the Regge ansatz 
according to Eqs.~(\ref{eq:rms1p},\ref{eq:rms1n}) on 
the Regge slope $\alpha_1^{'}$.  
For the quark distributions, 
the MRST01 NNLO parametrization \cite{Mar02} 
at scale $\mu^2$ = 1 GeV$^2$ was used in the calculations.
The shaded bands correspond to the experimental values.}
\label{fig:rmse}
\end{figure}

\begin{figure}[h]
\vspace{-1cm}
\epsfxsize=8.75cm
\leftline{\epsffile{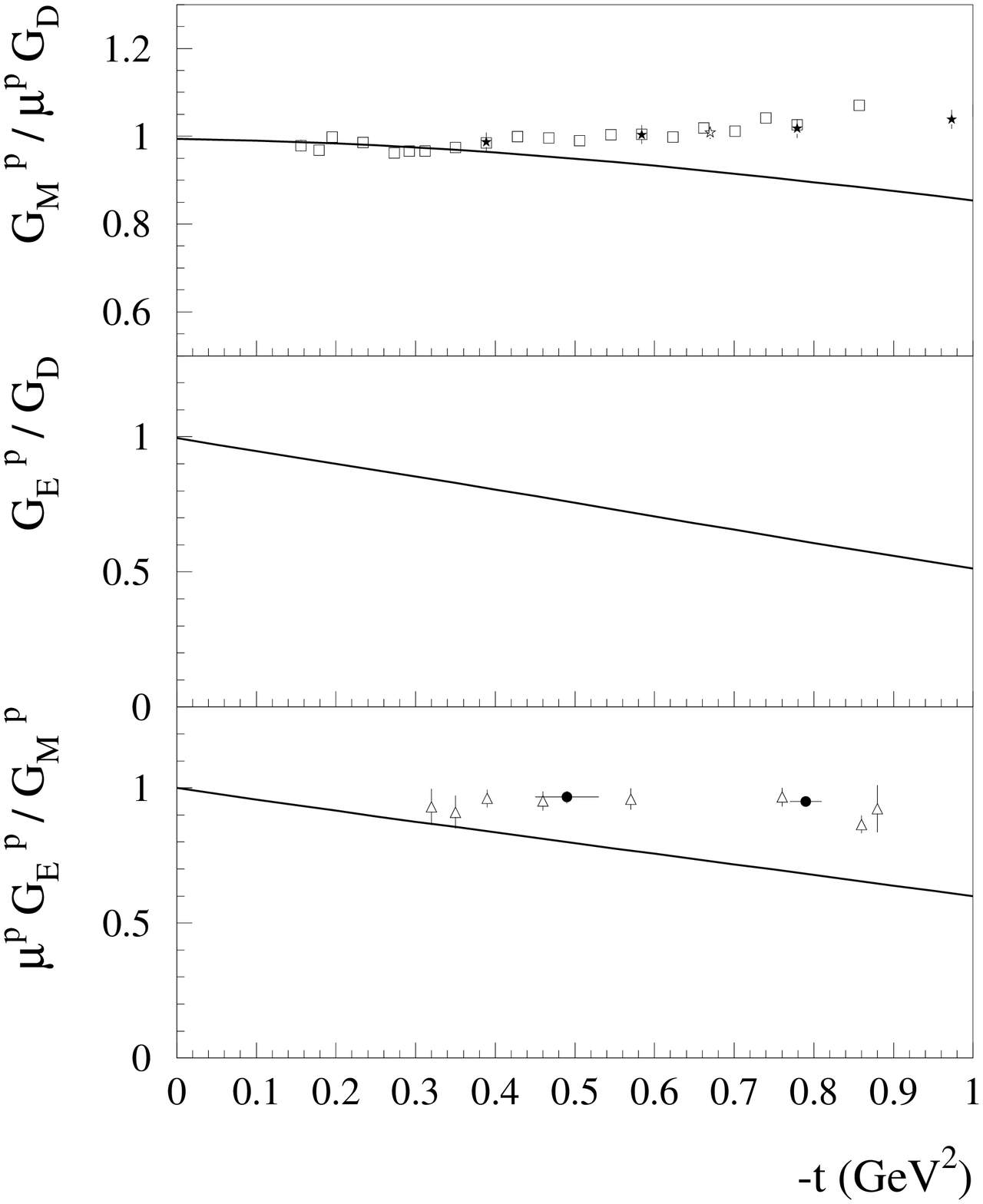}}
\vspace{-11.9cm}
\epsfxsize=8.75cm
\centerline{\hspace{8.8cm} \epsffile{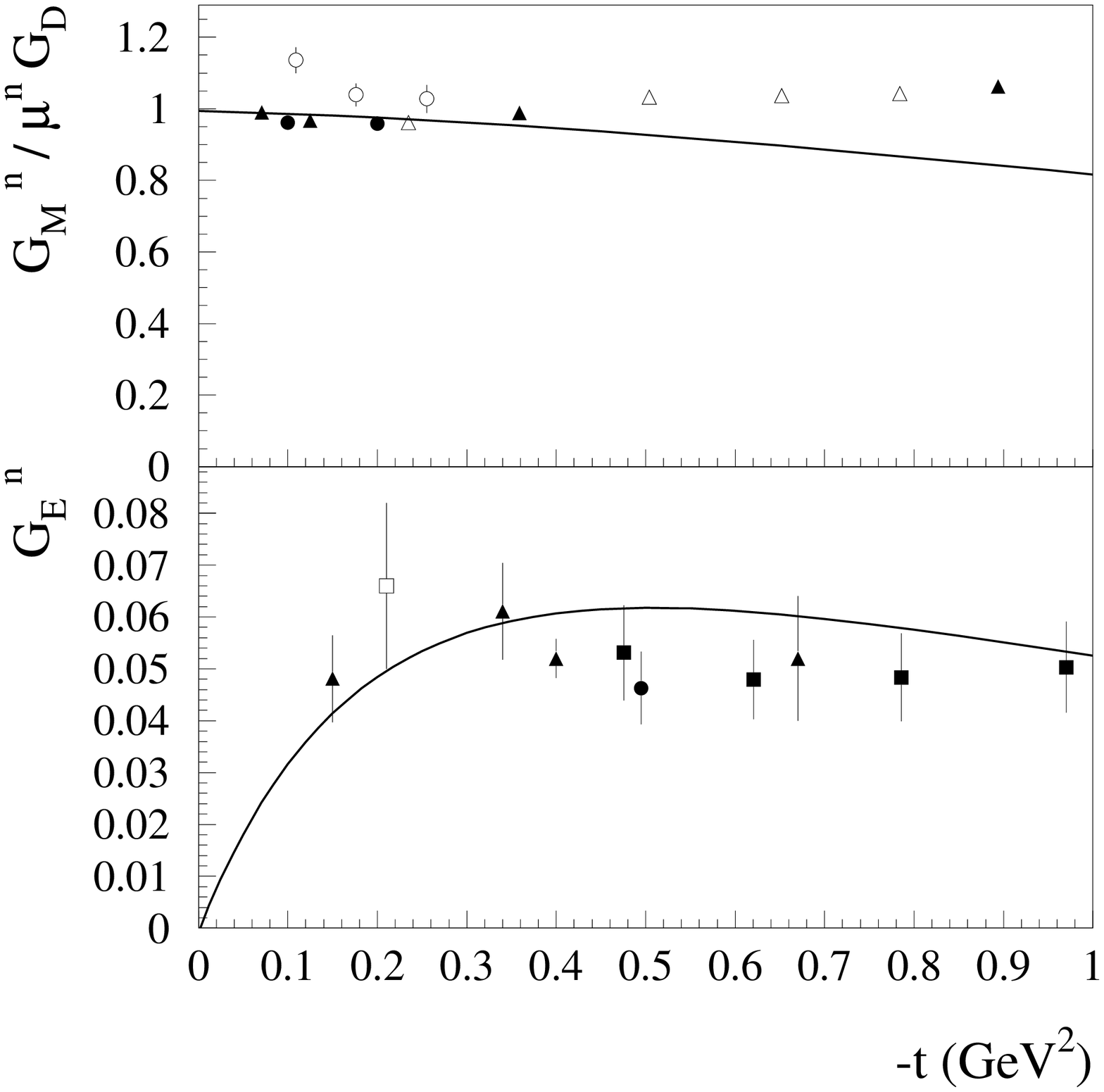}}
\vspace{.6cm}
\caption[]{\small Left side : 
proton magnetic (upper panel) and electric (middle panel) 
form factors compared to the dipole form 
$G_D(t) = 1/(1 - t/0.71)^2$, as well as 
the ratio of both form factors (lower panel). 
Right side : neutron 
magnetic (upper panel) and electric (lower panel) form factors.
The curves correspond to the Regge ansatz of 
Eqs.(\ref{eq:f1ud_1}) and (\ref{eq:f2ud_1}) , 
with $\alpha_1^{'} = 1.1$~GeV$^{-2}$, 
$\alpha_2^{'} = 1.1$~GeV$^{-2}$.  
The references to the data can be found in \cite{GPV02}.}
\label{fig:pn_ff}
\end{figure}

We start by discussing the 
$t$-dependence of the GPDs which is directly related to nucleon elastic
form factors through sum rules. In particular, 
the nucleon Dirac and Pauli form factors $F_1(t)$ and $F_2(t)$ 
can be calculated from the GPDs $H$ and $E$ through the 
following sum rules for each quark flavor ($q = u, d$) 
\begin{eqnarray}
F_{1}^{q}(t) \,=\, \int _{-1}^{+1}dx\; H^{q}(x,\xi ,t) \,  ,
\hspace{2cm} 
F_{2}^{q}(t) \,=\, \int _{-1}^{+1}dx\; E^{q}(x,\xi ,t) \, \; .
\label{eq:hesumrule} 
\end{eqnarray}
We can choose $\xi = 0$ in the previous equations, and  
model $H(x,0 ,t)$ and $E(x,0 ,t)$ subsequently.
For the GPD $H(x, 0, t)$, a plausible ansatz at low $-t$ is 
a Regge form as discussed in \cite{GPV01}.
This leads to the following
integrals to calculate the Dirac form factors for $u$- and $d$-quark flavors~:
\begin{eqnarray}
F_1^u(t) \,=\, \int _{0}^{+1}dx \; u_v(x) \;
{{1} \over {x^{\alpha_1^{\, '} t}}} \;,
\hspace{2cm}
F_1^d(t) \,=\, \int _{0}^{+1}dx \; d_v(x) \; 
{{1} \over {x^{\alpha_1^{\, '} t}}} \;,
\label{eq:f1ud_1}
\end{eqnarray}
where $u_v(x)$ and $d_v(x)$ are the $u$- and $d$-quark valence
distributions, and 
where $\alpha_1^{\, '}$ is the slope of the leading Regge trajectory. 
The proton and neutron Dirac form factors then follow from 
\begin{eqnarray}
F_1^p(t) \,=\, e_u \, F_1^u(t) \;+\; e_d \, F_1^d(t) \, , 
\hspace{2cm} 
F_1^n(t) \,=\, e_u \, F_1^d(t) \;+\; e_d \, F_1^u(t) \, , 
\label{eq:f1} 
\end{eqnarray}
where by construction $F_1^p(0)$ = 1, and $F_1^n(0)$ = 0.
\newline
\indent
Using the above ansatz, 
the Dirac mean squared radii of proton and neutron can be calculated as~:
\begin{eqnarray}
r^2_{1, p} &\,=\,& -6 \, \alpha_1^{\, '} \,
\int _{0}^{+1}dx \; 
\biggl\{ e_u \, \, u_v(x) \,+\, e_d \, \, d_v(x) \biggr\} \, \ln x \; ,
\label{eq:rms1p} \\
r^2_{1, n} &\,=\,& -6 \, \alpha_1^{\, '} \,
\int _{0}^{+1}dx \; 
\biggl\{ e_u \, \, d_v(x) \,+\, e_d \, \, u_v(x) \biggr\} \, \ln x  \;,
\label{eq:rms1n}
\end{eqnarray}
which yields for the electric mean squared radii 
of proton and neutron~:
\begin{eqnarray}
r^2_{E, p} \,=\, r^2_{1, p}
\;+\; {3 \over 2} \, {{\kappa_p} \over {m_N^2}} \; ,
\hspace{2cm}
r^2_{E, n} \,=\, r^2_{1, n}
\;+\; {3 \over 2} \, {{\kappa_n} \over {m_N^2}} \; ,
\label{eq:rmse}
\end{eqnarray}
where $\kappa_p$ ($\kappa_n$) are the proton (neutron) anomalous
magnetic moments. 
\newline
\indent
In Fig.~\ref{fig:rmse}, we show the proton and neutron rms radii as
function of the Regge slope $\alpha_1^{'}$, which is the only free
parameter in the ansatz of Eq.~(\ref{eq:f1ud_1}). 
One notes that the neutron rms radius is dominated by 
the Foldy term (term proportional to $\kappa_n$ in Eq.~(\ref{eq:rmse})), 
which gives $r^2_{E, n}$ = - 0.126 fm$^2$. Therefore, a relatively
wide range of values $\alpha_1^{'}$ are compatible with the neutron
data. However for the proton, a rather narrow range of values around 
$\alpha_1^{'} = 1.0 - 1.1$~GeV$^{-2}$ are favored. Such value is close
to the expectation from Regge slopes for meson trajectories, therefore
supporting the ansatz of Eq.~(\ref{eq:f1ud_1}).  
\newline
\indent
To calculate the electric and magnetic form factors of the nucleon,
one also needs to calculate the Pauli form factor $F_2$, besides $F_1$. 
For $F_2$, we use an ansatz which is 
based on a valence quark distribution for the valence part 
of $E(x, 0, t)$ entering in Eq.~(\ref{eq:hesumrule}). 
This leads to the following
integrals for the proton and neutron Pauli form factors~:
\begin{eqnarray}
F_2^u(t) \,=\, \int _{0}^{+1}dx \; 
\kappa_u {1 \over 2} \, u_v(x) \;
{{1} \over {x^{\alpha_2^{\, '} t}}} \; ,
\hspace{2cm}
F_2^d(t) \,=\, \int _{0}^{+1}dx \; 
\kappa_d \, d_v(x) \;
{{1} \over {x^{\alpha_2^{\, '} t}}} \;,
\label{eq:f2ud_1}
\end{eqnarray}
where $\kappa_u$ and $\kappa_d$ are given by 
$\kappa_u \,=\, 2 \, \kappa_p \,+\, \kappa_n$, and 
$\kappa_d \,=\, \kappa_p \,+\, 2 \, \kappa_n$.
\newline
\indent
In Fig.~\ref{fig:pn_ff},
we show the predictions of the above discussed Regge ansatz for the
proton and neutron form factors. For both proton
and neutron magnetic form factors, one sees that the Regge forms
reproduce the experimentally observed dipole behavior up to about 
$-t = 0.5$~GeV$^2$. Such behavior follows in the present ansatz from
the behavior of valence quark distributions at small/intermediate
values of $x$. At larger values of $-t$, the Regge form expectedly falls
short of the data as one expects a transition to the perturbative
behavior ($\sim 1/t^2$) of the magnetic form factors. 
For the ratio of electric to magnetic proton form factors, one interestingly
sees that the Regge form leads to a decreasing ratio with
$-t$. Although the simple Regge model falls too fast with $-t$, the
decreasing trend is in qualitative agreement with the data at larger $-t$ from
JLab \cite{Jon00,Gay02}. For the neutron magnetic form factors, one obtains
a remarkable good description up to $-t \simeq 1 $~GeV$^2$. 
\newline
\indent
The simple Regge ansatz discussed here \cite{GPV02}, catches the basic
features of the nucleon electromagnetic form factors at 
$-t < 0.5$~GeV$^2$. For $-t > 1$~GeV$^2$, an overlap
representation linking the nucleon Dirac form factor to GPDs has been
given in Refs.~\cite{Rady98,Die99}, describing the  trend of the
data. 
A topic for further study is to incorporate both small-$t$ and
large-$t$ regimes in a unified parametrization. This is needed to
perform the Fourier transform for the $t$-dependence of GPDs in order
to map out the distribution of partons in the transverse plane.

\section{DVCS beam-helicity asymmetry}

\begin{figure}[h]
\vspace{0cm}
\epsfysize=8.9cm
\leftline{\epsffile{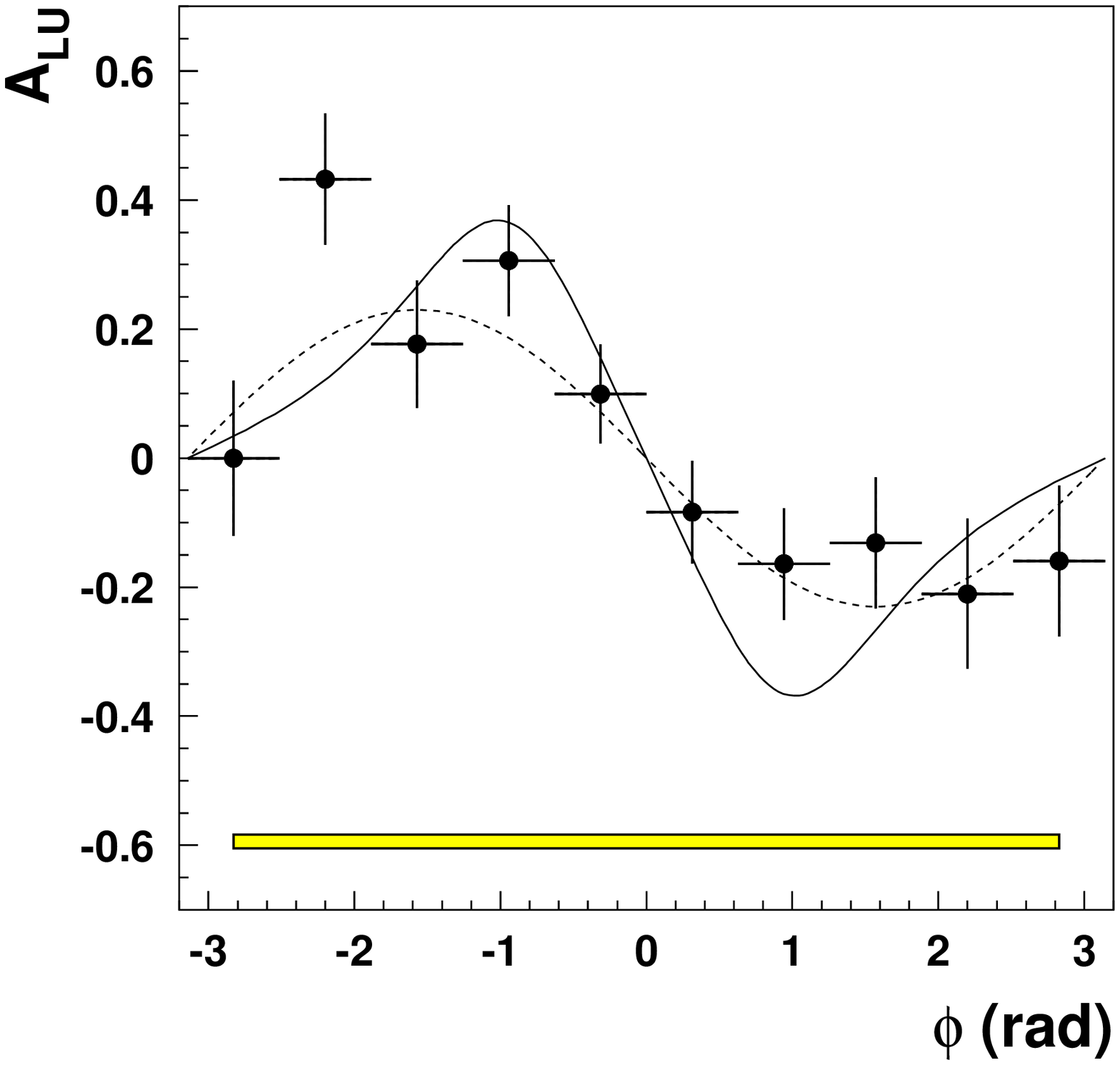}}
\vspace{-10.05cm}
\hspace{0.95cm}
\epsfysize=8.8cm
\rightline{\epsffile{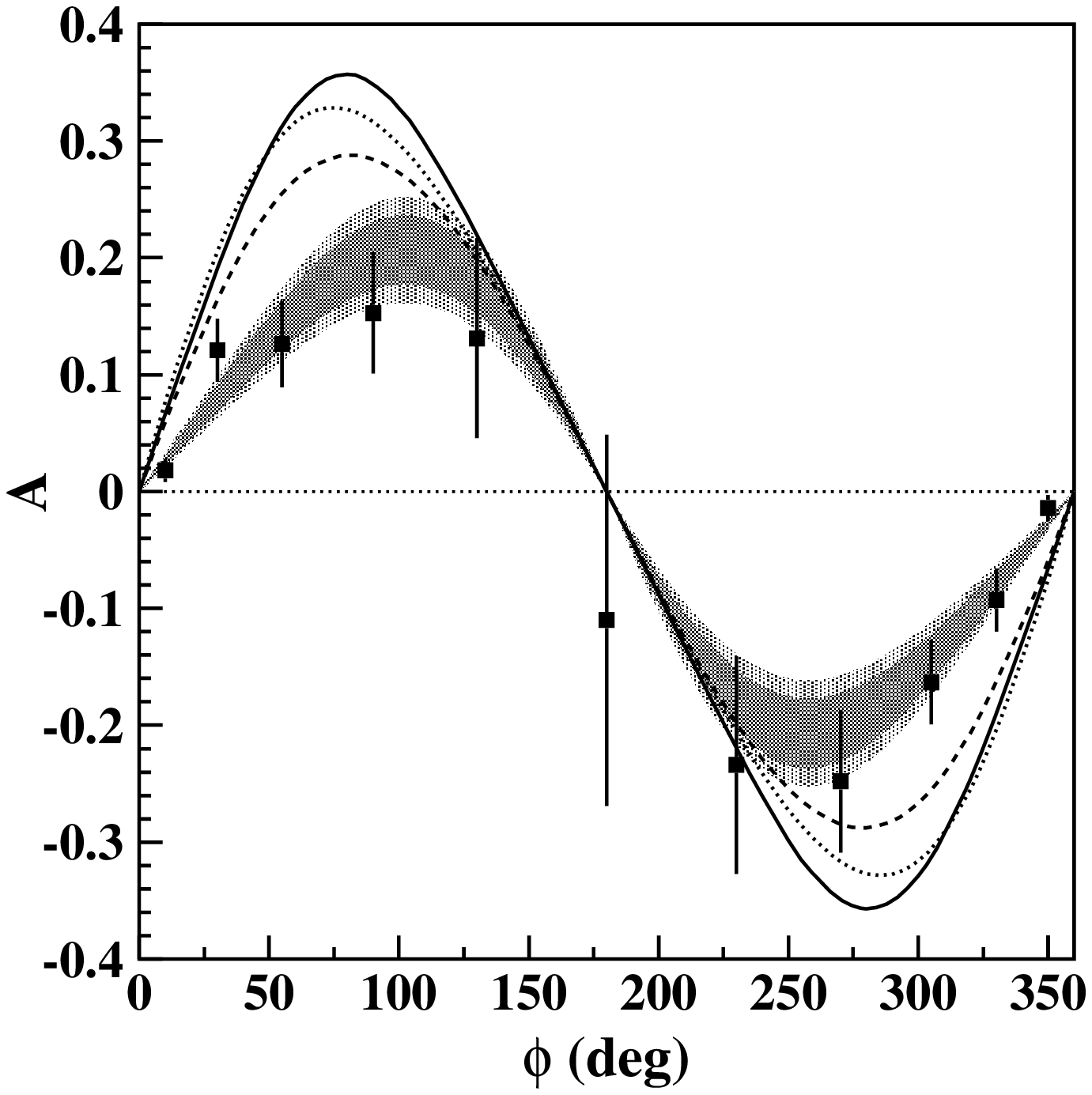}}
\vspace{-1.25cm}
\caption{The DVCS beam helicity asymmetry as measured at 
HERMES \cite{Air01} (left panel) and JLab/CLAS \cite{Step01} (right
panel). Full curves are the twist-2 + twist-3 predictions of 
Ref.~\cite{Kiv01}.}
\label{fig:ssaexp}
\end{figure}

We next turn to the DVCS observables and their dependence on the
GPDs. 
At intermediate lepton beam energies, one can extract the imaginary part
of the DVCS amplitude through the
$\vec e p \to e p \gamma$ reaction with a polarized lepton beam,
by measuring the out-of-plane angular dependence (in the angle $\phi$) 
of the produced photon \cite{Kro96}.
It was found in Refs.~\cite{Vdh98,Gui98} that the resulting electron
single spin asymmetry (SSA)
\begin{eqnarray}
{\cal A}^{\rm{SSA}} \;=\; {{\sigma_{e, h = +1/2} - \sigma_{e, h = -1/2}} \over
{\sigma_{e, h = +1/2} + \sigma_{e, h = -1/2}}} \, ,
\label{eq:ssa}
\end{eqnarray}
with $\sigma_{e, h}$ the cross section for an electron of helicity $h$,
can be sizeable for HERMES ($E_e$ = 27 GeV) and JLab ($E_e$ = 4 - 11 GeV)
beam energies.
The SSA for the $\vec e p \to e p \gamma$ 
reaction has recently been measured in pioneering experiments at
HERMES \cite{Air01} and JLab/CLAS \cite{Step01}, which are shown in 
Fig.~\ref{fig:ssaexp}.
These experiments display already at the accessible values of $Q^2
\simeq 1 - 2.5 $~GeV$^2$ predominantly a $\sin \phi$ dependence, 
indicating a dominance of the twist-2 DVCS amplitude. Furthermore,
the observed magnitude is in good agreement with the theoretical
calculations of Refs.~\cite{Kiv01,Bel02} in terms of GPDs.
These first experiments give strong indication that the SSA is a promising
observable to get access to GPDs. Once the leading order mechanism in
terms of GPDs is confirmed by experiment, the measured helicity
difference in Eq.~(\ref{eq:ssa}) is 
directly proportional to the GPDs along the line $x = \xi$, and one
may proceed to map out the 'envelope' function $H(\xi, \xi, t)$,
and analogously for $E$, $\tilde H$ and $\tilde E$.
\newline
\indent
Dedicated experiments to measure the SSA with improved
accuracy in a large kinematic range are already planned and underway
both at JLab \cite{Elo02,Mec02} and HERMES \cite{Has02,Rit02}.

\section{DVCS beam-charge asymmetry}

\begin{figure}[h]
\vspace{0cm}
\epsfxsize=12.cm
\leftline{\hspace{-2cm} \epsffile{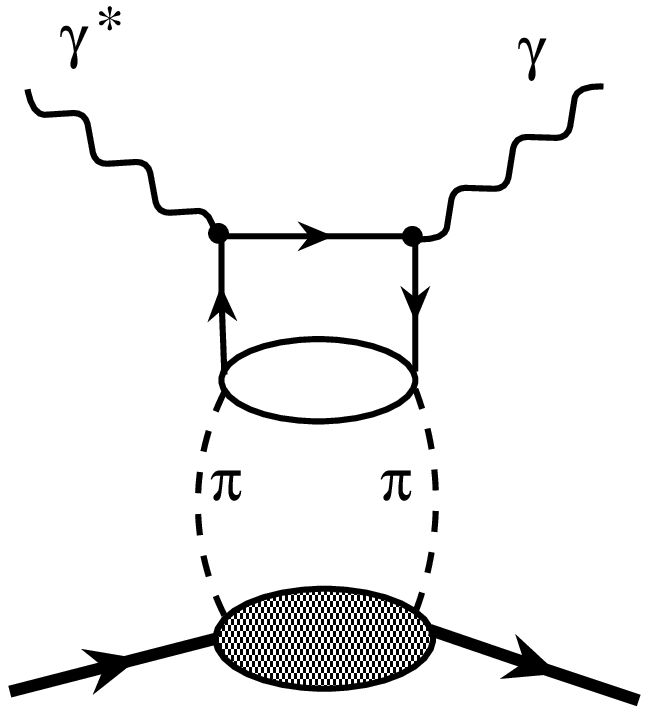}}
\vspace{-10.cm}
\epsfxsize=10cm
\centerline{\hspace{6.5cm}\epsffile{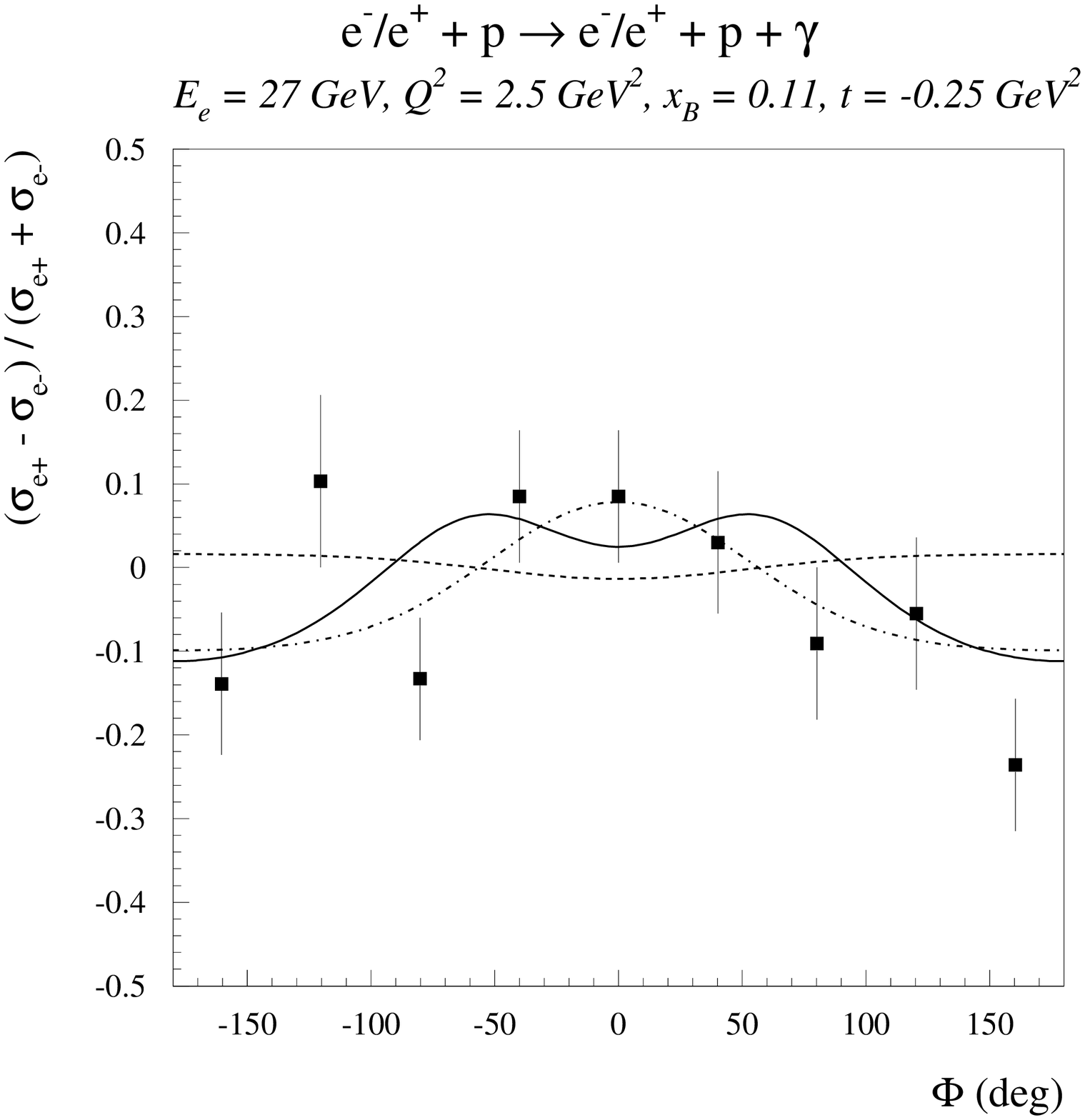}}
\vspace{-1cm}
\caption{Left side : Model contribution to the D-term entering the
GPDs $H$ and $E$.
Right side : the DVCS beam-charge asymmetry. 
The preliminary data are from HERMES \cite{Ell02}. 
Theoretical predictions are from Ref.~\cite{Kiv01} :
the calculations of the twist-2 DVCS amplitude with (without) D-term
are shown by the dashed-dotted (dashed) curves. The full curve is
obtained when adding twist-3 effects in addition to the D-term.}
\label{fig:chasy}
\end{figure}

Besides the beam-helicity asymmetry for the  $\vec e p \to e p \gamma$ 
reaction, which accesses the imaginary part of the DVCS amplitude, 
one gets access to the real part of the DVCS amplitude by measuring
both $e^+ p \to e^+ p \gamma$ and  $e^- p \to e^- p \gamma$ processes.
In those reactions, besides the mechanism where 
the photon originates from a quark (handbag diagrams of
Fig.~\ref{fig:handbags}, the photon can also be emitted by the lepton
lines, in the so-called Bethe-Heitler (BH) process. 
Because the BH amplitude contains two lepton electromagnetic couplings
in contrast to the DVCS process, the interference between BH and
DVCS processes changes sign when comparing the $e^+ p \to e^+ p \gamma$ and
 $e^- p \to e^- p \gamma$ reactions. Therefore, in the difference of
cross sections $\sigma_{e^+} - \sigma_{e^-}$, the BH drops out, and
one measures the real part of the BH-DVCS interference \cite{Bro72}
\begin{equation}
\sigma_{e^+} - \sigma_{e^-} \sim  \Re  \left[ T^{BH} {T^{DVCS}}^* \right] \;,
\end{equation}
which is proportional to the {\it real} 
(principle value integral) part of the DVCS amplitude. 
In this way, the difference $\sigma_{e^+} - \sigma_{e^-}$
is sensitive to the GPDs away from the line $x = \xi$.
\newline
\indent
It has been shown in \cite{Kiv01} that this beam-charge asymmetry
gets a sizeable contribution from the D-term. The latter encodes 
$q \bar q$ scalar-isoscalar correlations in the nucleon as shown
in Fig.~\ref{fig:chasy} (left side), and has been estimated in the chiral
quark soliton model \cite{Pet98}.
\newline
\indent
The beam-charge asymmetry associated with DVCS has been accessed
experimentally very recently at HERMES, as reported in this Workshop
\cite{Ell02}. The preliminary data are shown in Fig.~\ref{fig:chasy},
together with the theoretical predictions. The experimental asymmetry
shows a $\cos \phi$ dependence with magnitude $\sim 0.10 - 0.15$, and
favors the calculations which include the D-term. This opens up the
perspective to study (mesonic) $q \bar q$ components in the nucleon 
wavefunction in a systematic and model independent way. 
Further measurements of the beam-charge asymmetry associated with DVCS
with improved statistics are planned at HERMES \cite{Rit02}.

\section{Hard meson electroproduction (HMP)}

\begin{figure}[h]
\vspace{-1.5cm}
\epsfxsize=9.75cm
\centerline{\epsffile{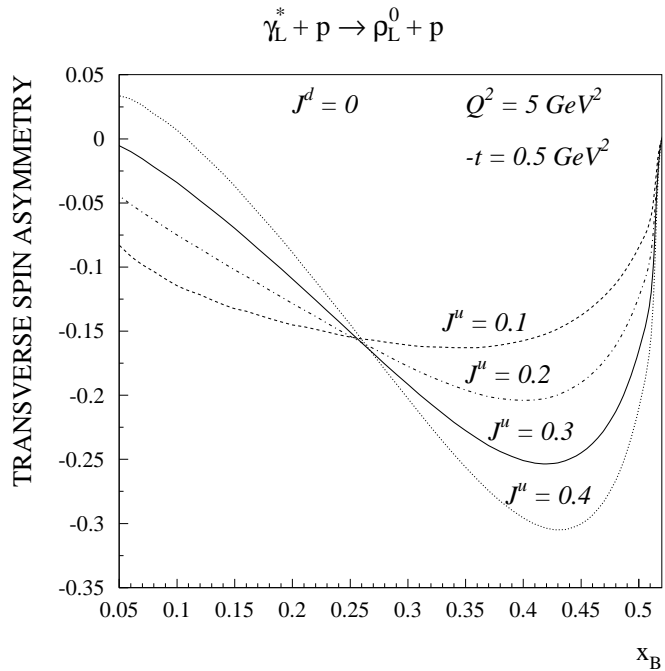}}
\vspace{-1cm}
\caption{\small $x_B$ dependence of the transverse spin asymmetry for the 
$\gamma^*_L \vec p \to \rho_L^0 p$ reaction. 
The estimates are given using the model for the GPDs $E^u$ and
$E^d$ as described in Ref.~\cite{GPV01}.
The curves show the sensitivity to the value of $J^u$ 
as indicated on the curves (for a value $J^d = 0$).} 
\label{fig:rhoo_asy_0p5}
\end{figure}

The GPDs reflect the structure of the nucleon 
independently of the reaction
which probes the nucleon. In this sense, they are universal quantities and can
also be accessed, in different flavor combinations, through the hard exclusive
electroproduction of mesons 
- \( \pi ^{0,\pm },\eta ,...,\rho ^{0,\pm },\omega ,\phi ,... \) - 
for which a QCD factorization proof was given
in Ref.~\cite{Col97}. This factorization theorem 
applies when the virtual photon is longitudinally 
polarized, which corresponds to a small size configuration 
compared to a transversely polarized photon. 
\newline
\indent
In the following, we consider the vector meson electroproduction processes 
$\gamma ^{*}_{L} + N \rightarrow V_L + N$ at large $Q^2$ on the nucleon $N$, 
where $V_L$ ($\rho^0_L$, $\rho^+_L$, $\omega_L$, ...) 
denotes the produced vector meson with longitudinal polarization. 
For $\rho^0_L \, p$ electroproduction on the proton, 
the leading order amplitudes involving no spin-flip of the nucleon
($A$) or involving a nucleon spin-flip ($B$) are proportional to the
following combinations of GPDs~\cite{Man98,Vdh98}~:
\begin{eqnarray}
A_{\rho^0_L \, p} \,&=&\, \int_{-1}^1 dx \; 
{1 \over {\sqrt 2}} \, {\left(e_u \ H^u \,-\, e_d \ H^d\right)} 
\; \left\{ {{1} \over {x - \xi + i \epsilon}}
+ {{1} \over {x + \xi - i \epsilon}} \right\} , 
\label{eq:arhoo} \\
B_{\rho^0_L \, p} \,&=&\, \int_{-1}^1 dx \;
{1 \over {\sqrt 2}} \, {\left(e_u \ E^u \,-\, e_d \ E^d\right)} 
\;\left\{ {{1} \over {x - \xi + i \epsilon}}
+ {{1} \over {x + \xi - i \epsilon}} \right\} ,
\label{eq:brhoo}
\end{eqnarray}
where $e_u = +2/3$  ($e_d = -1/3$) are the $u$ ($d$) quark 
charges respectively.  
\newline
\indent
Recently, $\rho^0_L$ data have been obtained at HERMES 
for $Q^2$ up to 5 GeV$^2$, around 
c.m. energy $W \approx$ 5 GeV \cite{Air00}. 
The calculations for $\rho^0_L$, including a model
estimate for power corrections \cite{Vdh99},   
point towards the dominance of the quark exchange mechanism 
in this intermediate $W$ range.
However due to the large size of the power corrections at accessible
values of $Q^2$, it would be too premature at the present stage 
to try to extract quark GPDs from these 
vector meson electroproduction cross sections. 
To reach this goal, one first 
needs to get a better theoretical control over 
the power (higher-twist) corrections, which is an important topic for  
future work. 
\newline
\indent
Besides the cross section $\sigma_L$, the second
leading order observable for HMP, 
is the transverse spin asymmetry, ${\cal A}_{V_L N}$ 
(TSA) for a proton target polarized perpendicular to the reaction plane. 
For the electroproduction of longitudinally polarized vector mesons, 
induced by a longitudinal virtual photon, 
the TSA is given by \cite{GPV01}~: 
\begin{equation}                                         
{\cal A}_{V_L N} = - \, {{2 \, |\Delta_\perp|} \over {\pi}} \,
\frac{{\rm{Im}} (A B^*) / m_N}
{|A|^2 \, (1-\xi^2) - |B|^2 \, \left(\xi^2 + t / (4 m_N^2) \right) 
- {\rm{Re}}(AB^*) \, 2 \, \xi^2} \, ,
\label{eq:rhopasy}
\end{equation}
which is proportional to the modulus 
$|\Delta_\perp|$ of the perpendicular component of the momentum transfer 
$\Delta$, and 
with $A$ and $B$ given by Eqs.~(\ref{eq:arhoo}-\ref{eq:brhoo}) 
in the case of $\rho^0_L$. 
One sees that the TSA is proportional to the 
imaginary part of the {\it interference} of the amplitudes $A$ and $B$, 
which contain the GPDs $H$ and 
$E$ respectively. Therefore, it depends {\it linearly} on the GPD $E$. 
Note that in contrast, both in the DVCS cross sections and SSA as well as 
in the cross sections for (longitudinally polarized) vector mesons, 
the GPD $E$ only enters besides a large contribution of the GPD $H$. 
Therefore, the transverse spin asymmetry of Eq.~(\ref{eq:rhopasy}) provides   
a unique observable to extract the GPD $E$.  
\newline
\indent
Besides, one may expect that 
the theoretical uncertainties and open questions for the 
meson electroproduction cross sections largely disappear for the 
TSA, suggesting that  
it is less sensitive to pre-asymptotic effects and 
that the leading order expression of Eq.~(\ref{eq:rhopasy}) 
is already accurate  
at accessible values of $Q^2$ (in the range of a few GeV$^2$).
Due to its linear dependence on the GPD $E$, the TSA  
for longitudinally polarized vector mesons opens up the perspective to 
extract the total angular momentum contributions $J^u$ and 
$J^d$ of the $u-$ and $d$-quarks to the proton spin. 
In the parametrization for the GPDs $E^q$ presented in Ref.~\cite{GPV01},
$J^u$ and $J^d$, enter as free parameters.
Due to the different $u$- and $d$-quark content of the vector mesons, 
the asymmetries for the $\rho^0_L$, $\omega_L$ and $\rho^+_L$ channels 
are sensitive to different combinations of $J^u$ and $J^d$,  with
$\rho^0_L$ production mainly sensitive to the combination $2 J^u + J^d$, 
$\omega_L$ to the combination $2 J^u - J^d$, and 
$\rho^+_L$ to the isovector combination $J^u - J^d$.
\newline
\indent
In Fig.~\ref{fig:rhoo_asy_0p5}, the sensitivity of the TSA for 
$\rho^0_L$ production on different values of $J^u$ is shown. 
The TSA for $\rho^0_L$ electroproduction displays 
a pronounced sensitivity to $J^u$ around 
$x_B \approx 0.4$, where asymmetries are predicted in the -15 \% to -30 \% 
range according to the value of $J^u$. 
It will therefore be very 
interesting to provide a first measurement of this asymmetry 
in the near future, for a transversely polarized target, such as is
currently available at HERMES.
\newline
\indent
In conclusion, we have seen some very promising first glimpses of GPDs 
entering hard exclusive reactions at the existing facilities. 
A dedicated program aiming at the extraction of the full physics
potential contained in the GPDs will also require a dedicated facility 
combining high luminosity and a good resolution 
(in order to fully resolve the exclusive final state),
as discussed at this Workshop, see~\cite{Gui02,Har02,Mil02}.

\end{document}